\documentclass[12pt]{iopart}

\usepackage{iopams}

\usepackage{graphicx}
\usepackage{color}
\usepackage{dcolumn}
\usepackage{bm}
\usepackage{amssymb}
\usepackage{appendix}

\newcommand{\Ham}[0]{\mathcal{H}}
\newcommand{\ims}[0]{\mbox{\scriptsize i}}
\newcommand{\scalarprod}[2]{\langle #1 | #2 \rangle}
\begin{document}
 
\title[Quantum dot embedded in AB ring]{Time-dependent wave packet simulations
  of transport through Aharanov-Bohm rings with an embedded quantum dot}

\author{C.\ Kreisbeck$^{1}$, T.\ Kramer$^{2,3}$, R.A.\ Molina$^{4}$}

\address{ $^{1}$  Department of Chemistry and Chemical Biology, Harvard University, Cambridge, Massachusetts 02138, USA\\
$^{2}$ Zuse Institute Berlin, Berlin, Germany\\
$^{3}$ Department of Physics, Harvard University, Cambridge, Massachusetts 02138, USA \\
$^{4}$ Instituto de Estructura de la Materia - CSIC, 28006 Madrid, Spain}

\begin{abstract}
We have performed time-dependent wave packet simulations of realistic Aharonov-Bohm (AB) devices with a quantum dot embedded in one of the
arms of the interferometer. The AB ring can function as a measurement device for the intrinsic transmission phase through the quantum dot,
however, care has to be taken in analyzing the influence of scattering processes in the junctions of the interferometer arms. We consider a
harmonic quantum dot and show how the Darwin-Fock spectrum emerges as a unique pattern in the interference fringes of the AB oscillations.
\end{abstract}


\section{Introduction}

Interference effects are fundamental for the description of transport in mesoscopic devices. Transport properties of a quantum
mechanical system are characterize by the complex transmission coefficient $t$. The modulus of the transmission coefficient is
readily available experimentally by measuring the conductance in a two-terminal configuration. According to the Landauer-B\"uttiker picture,
the conductance for mono-channel leads is proportional to the transmission
probability through the system $g=(e^2/h)|t|^2$ \cite{Landauer70,Buttiker86}.
The phase of the transmission coefficient $\alpha$ ($t=|t|e^{i\alpha}$) is, however, a more elusive quantity for experiments.

Phase sensitive experiments were pioneered by Yacoby {\em et al.} in the 90s
\cite{Yacoby95}. A Quantum Dot (QD) was embedded in one arm of an Aharanov-Bohm (AB) ring. 
Oscillations in the conductance of the device as a function of the magnetic
flux were followed while varying a plunger gate voltage $V_g$ affecting the dot.
The relative phase between the pattern of oscillations for different values of
$V_g$ would, in principle, depend on the transmission phase of the quantum dot.
The device, however, was a two-terminal setup and Onsager relations dictate
that the dependence of the conductance of the whole device has to be an even
or odd function of the magnetic flux. Then, the relative phases between the
oscillations at different values of $V_g$ can only be $0$ or $\pi$ \cite{Buttiker86,Onsager31}.
 
A multiple terminal configuration lifts this phase locking condition. Multiterminal devices for extracting 
the phase information were fabricated by Schuster {\em at al.} and transmission phases could be extracted
from the measured data \cite{Schuster97}.
The expected Breit-Wigner behavior of the measured transmission phase in each resonance was found, giving credit to the claim
of measuring the intrinsic transmission phase of the embedded QD through the analysis of the interference patterns
in the conductance of the whole AB device.
However, more unexpectedly, the behavior of the phase was the same between all measured resonances. 
Transmission zeros between every pair of neighboring resonances and the
associated $\pi$ phase lapses were responsible for the in-phase behavior of the sequences of resonances.

This seminal series of works motivated further experiments and a great wealth of theoretical works. 
On the experimental side, the role of the magnetic field was explored by Sigrist {\em at al.} in AB ring with
one QD embedded in each of its arms \cite{Sigrist04}. Phase lapses were found for only specific ranges of values of
the magnetic field. The phase of a QD in the Kondo regime has also received a lot of attention after Gerland {\em et al.  }
predicted that the transmission phase of a QD measured in the AB device should evolve 
between the $\pi$ results of Coulomb blockade to the $\pi/2$
result predicted in the Kondo regime depending on temperature
and width \cite{Gerland00}. The first experiments in the Kondo regime reported a $3\pi/2$ phase shift along the spin-degenerate pair, a surprising result that is still not properly explained \cite{Ji00,Ji02}. Further experiments claimed to have observed the $\pi/2$ phase shift corresponding to the Kondo regime but the temperature of this experiments was much higher than the Kondo temperature and outside the regime where the Kondo result is expected to hold\cite{Zaffalon08}. 
More recent experiments by Takada {\em et al.} have managed to measure the transmission phase shift through a Kondo correlated quantum dot and find the transition between the Kondo regime and Coulomb blockade regime with excellent agreement with Numerical Renormalization Group calculations \cite{Takada14}. Effects of the ratio between width and interaction strength have also been experimentally investigated \cite{Takada16}.
Avinun-Kalish {\em at al.} investigated
smaller dots and found the crossover from the universal regime where phase
lapses occur in between every neighboring resonance to a mesoscopic regime
where phase lapses occur in a random fashion was observed
when decreasing the number of electrons from $20$ down to $0$ \cite{Avinun05}.

As it became clear, thanks to the work by Levy-Yeyati and B\"uttiker, phase
lapses were associated with zeros in the transmission 
coefficient, the only situation in which the phase of the complex number as a
function of the energy $E$ or as a function of $V_g$ can have a discontinuity
\cite{LevyYeyati95,Lee99,Taniguchi99}. In non-interacting models, zeros of the transmission in the valley between two peaks
would appear in between consecutive resonances of the same parity with respect
to the lead position\cite{LevyYeyati00}.
Most of the theoretical works were devoted to the objective of
explaining the universal sequences of resonances and phase lapses appearing in experiments 
\cite{Hackenbroich97,Baltin99a,Silvestrov00,Silvestrov07,Goldstein09, Karrasch07,Silva02,Oreg07,Molina12}.
Several works have addressed the importance of the electronic correlations in the transition to
the universal regime \cite{Karrasch07}. However, many-body numerical calculations have challenged
the interpretation of these results \cite{Molina13}. 
Molina {\em et al.} found that wavefunction correlations in chaotic ballistic
quantum dots could be responsible for the crossover between the mesoscopic regime
and universal regime \cite{Molina12}. A more detailed statistical analysis have investigated the stability 
of the chaotic correlations under the influence of fluctuations \cite{Jalabert14}. New experiments with a more systematic statistical analysis of the results are needed to discriminate between different theoretical approaches trying to explain the experimental transition between the mesoscopic and the universal regimes. 

A different range of theoretical works have tried to analyze up to what extend the experimental results
can be trusted to give unambiguous information about the intrinsic transmission phase of the quantum dot.
Taking into account
the different harmonics due to multiple turns around the AB ring the
conductance of the whole device can be written as
\begin{equation}
g=g_0+\sum_n g_g \cos{\left(2\pi n \phi / \phi_0 + \beta_n \right)},
\label{eq:AB}
\end{equation}
with the quantum flux $\phi_0$. 
The connection between the measured phase of the conductance oscillations $\beta_1$ and the transmission phase
$\alpha$ was discussed in several works \cite{Taniguchi99,Wu98,Kang99,Aharony02,EntinWohlman02}.
The conclusion that could be extracted from
these series of works was that under not too restrictive constraints
regarding the coupling to the environment the transmission phase 
$\alpha$ can be extracted from $\beta_1$ in multiterminal
geometries. However, the main focus was in the opening of the systems and in partial coherent transport using one-dimensional models
for the leads without taking into account other distorting effects like possible scattering in the junctions of the interferometric
arms. Experimental advances in this direction have been made recently trying to find more clear criteria about when the phase measurements can be trusted and when the contributions from multiple path can be neglected\cite{Takada15}. 

A different approach was taken by Fischer {\emph et al.} \cite{Buchholz2009a}. 
The purpose was to make very controllable experiments that could be verified by theoretical
simulations. They wanted to understand the behavior of the complete device
taking into account the two-dimensional character of waveguides and
scattering effects due to the junctions and in the leads \cite{Kreisbeck10}.
The experiments and theoretical calculations in pure AB rings demonstrated the strong influence
of scattering effects in the junctions and arms of the ring that were by
themselves the sources of phase jumps in the pattern of conductance
oscillations as a function of the magnetic field.

The aim of this work is to study the functioning of a realistic four terminal
AB device as a phase detector of an embedded quantum dot. For this purpose we use time-dependent wave packet
simulations \cite{Kreisbeck10,Kramer10} of the whole device from which we extract the complex
transmission at different energies from a single run. For our simulations we use a variation of
the device fabricated by Fischer {\emph et al.} \cite{Buchholz2009a} including a harmonic dot in one of
the arms of the interferometer. Using this configuration has several
advantages: it has been shown to be experimentally feasible and the pure interferometer without dot
has been theoretically modeled before \cite{Kreisbeck10,Kramer16}.
Due to the size
of the dot we take into account the full effect of the magnetic 
field and cyclotron orbits inside the dot. Although, scattering effects in the
cross-junctions of the device can distort the measurement, careful simulations
of the device without dot allow us to predict scattering free regions where
the intrinsic dot transmission phase can be extracted from the
measurements. The simulations show Zeeman cat's whiskers appearing in
the AB oscillation pattern of the device as a signature of the Darwin-Fock
spectrum of the dot. As resonances in a Zeeman multiplet have the same parity, phase
lapses appear in between all resonances which allow us to identify the quantum
numbers corresponding to each resonant peak and use the device for
simultaneous phase detection and transport spectroscopy.

The remainder of the manuscript is organized as follows. In Section~2 we outline the used theoretical model and methodology. We summarize the transmission properties of the embedded quantum dot defined by the Darwin-Fock spectrum in Section~3.
After that, we demonstrate in Section~4 that the AB-ring geometry can be configured as transmission phase detector of an embedded quantum dot. We show that the Darwin-Fock
spectrum emerges as unique patterns in the AB interference fringes.

\section{Methodology}

\begin{figure}[t!]
\begin{center}
\includegraphics[width=0.7\columnwidth]{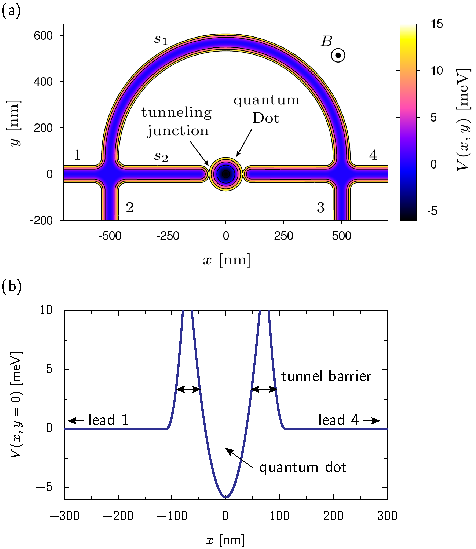}
\caption{\label{fig:PotentialABRingembeddedQDot}
(a) Potential landscape of the AB interferometer configured as a transmission phase detector
of an embedded quantum dot. The ring is constructed in the same way as in \cite{Kreisbeck10}.
We assume an isotropic harmonic quantum dot.
(b) shows a cut in $x$-direction of the potential along the center of the lower arm $s_2$ of the ring ($y$=0). We configure the quantum dot into the
resonant tunneling regime.
The distance between the caps and the center
of the quantum dot adjusts the tunneling coupling. We use a distance of 110~nm.
}
\end{center}
\end{figure}

A sketch of the setup including information on the parameters used in our
simulations is given in the top panel of Fig.~\ref{fig:PotentialABRingembeddedQDot}.
The design is based on the experimental device without QD used in
Ref.~\cite{Kreisbeck10}. This allow us to easily compare between
results for the device without QD and for the device with QD and to guide the
design and fabrication of a real device. A similar design has also been used for mode filtering
\cite{Buchholz11} which can be necessary for a clean measurement of the transmission
phase. The experimental rings were fabricated in AlGaAs/GaAs
heterostructures. The mean-free path $\ell_e\sim8\--15$~$\mu$m is larger than
the size of the device geometry. The waveguide had 
radii of $0.5\--1$~$\mu$m. This ensures coherent
electron transport along the ring and minimizes scattering at impurities and inelastic scattering.
Similar experimental devices and theoretical techniques have been used to study thermal currents \cite{Kramer16}.

We use the effective mass approximation and describe the electron
transport in terms of a single particle picture. The effective mass for the AlGaAs/GaAs heterostructure is given by $m^\ast=0.067m_e$, where
$m_e$ stands for the electron mass.
The magnetic field is included by minimal coupling
replacing $\mathbf{p}\rightarrow-i\hbar\nabla+e \textbf{A}(x,y)$.
We describe the electron
transport within the ballistic regime and interpret the experimental outcome
in terms of elastic scattering. We compute
the transmission amplitudes with a time-dependent 
wave-packet approach. The basic ideas for the application of
the wave-packet approach to mesoscopic physics are given in
Ref.~\cite{Kramer08,Chaves09,Kramer10}.
The time-dependent approach is based on wave-packet propagation. While the transversal
shape of the wave packets is governed by the harmonic confinement of the asymptotic leads,
we are free to choose the longitudinal component. In the following, we
use Gaussian wave packets. 
The time-dependent approach has several advantages:
(i) the quantum choreography of the propagated pulse gives an
intuitive physical picture of the scattering mechanisms involved in the transfer
process, (ii) the wave packet propagation efficiently computes the energy resolved
transmission amplitudes for a large range of energies, and (iii) a time-dependent approach allows
us to simulate general multi-terminal devices with complicated geometries and topologies.
More details of the simulation and snapshots of the time evolution can be
found in the Appendix. 

In order to perform realistic simulations we incorporate the 2d potential
landscape (see Fig.~\ref{fig:PotentialABRingembeddedQDot}). Although the whole
procedure is within a non-interacting particle picture, the potential
landscape in the leads and in the quantum dot effectively take into account depletion effects due to Coulomb interaction. 
Other effects of the Coulomb interaction like the charging energy of the dot and effects of strong correlations beyond mean field
are not taken into account in our approach. We expect charging effects in the dot to separate the levels in energy essentially without modifying the phase information in the measurements.
The asymptotic leads are given as harmonic waveguides with a confining frequency
$\hbar\omega=5$~meV. 
We accurately model the junctions as rounded orthogonal cross-junctions ($R_{\rm junction}=70$~nm). 
The latter induce collimation effects that significantly change the scattering
behavior \cite{Baranger91}.
We set the radius of the ring to $R=0.5$~$\mu$m. 
To avoid phase locking due to device symmetries \cite{Kreisbeck10}, we configure the four-terminal device in the non-local setup, in which
voltage and current probes are spatially separated. 
In this configuration the current $I_{43}$ flows between lead~3 and lead~4, while the voltage drop $V_{12}$ is
detected at lead~1 and lead~2.
We define the non-local resistance
\begin{equation}\label{eq:expsetup1}
R_{43,12}=\frac{V_{12}}{I_{43}}.
\end{equation}

To avoid disturbing multi-mode effects, we configure the interferometer into a transport regime
where only two open modes contribute to the transport.
A mode-filtered electron injection allows experimentalists to perform even
single-mode transport in a multi-mode waveguide structure \cite{Buchholz11}.
Visualization of the transmission phase of the
embedded quantum dot requires a continuous phase drift of the AB oscillations while
scanning through the Fermi energy \cite{Avinun05}. 
\begin{figure}[t!]
\begin{center}
\includegraphics[width=0.99\columnwidth]{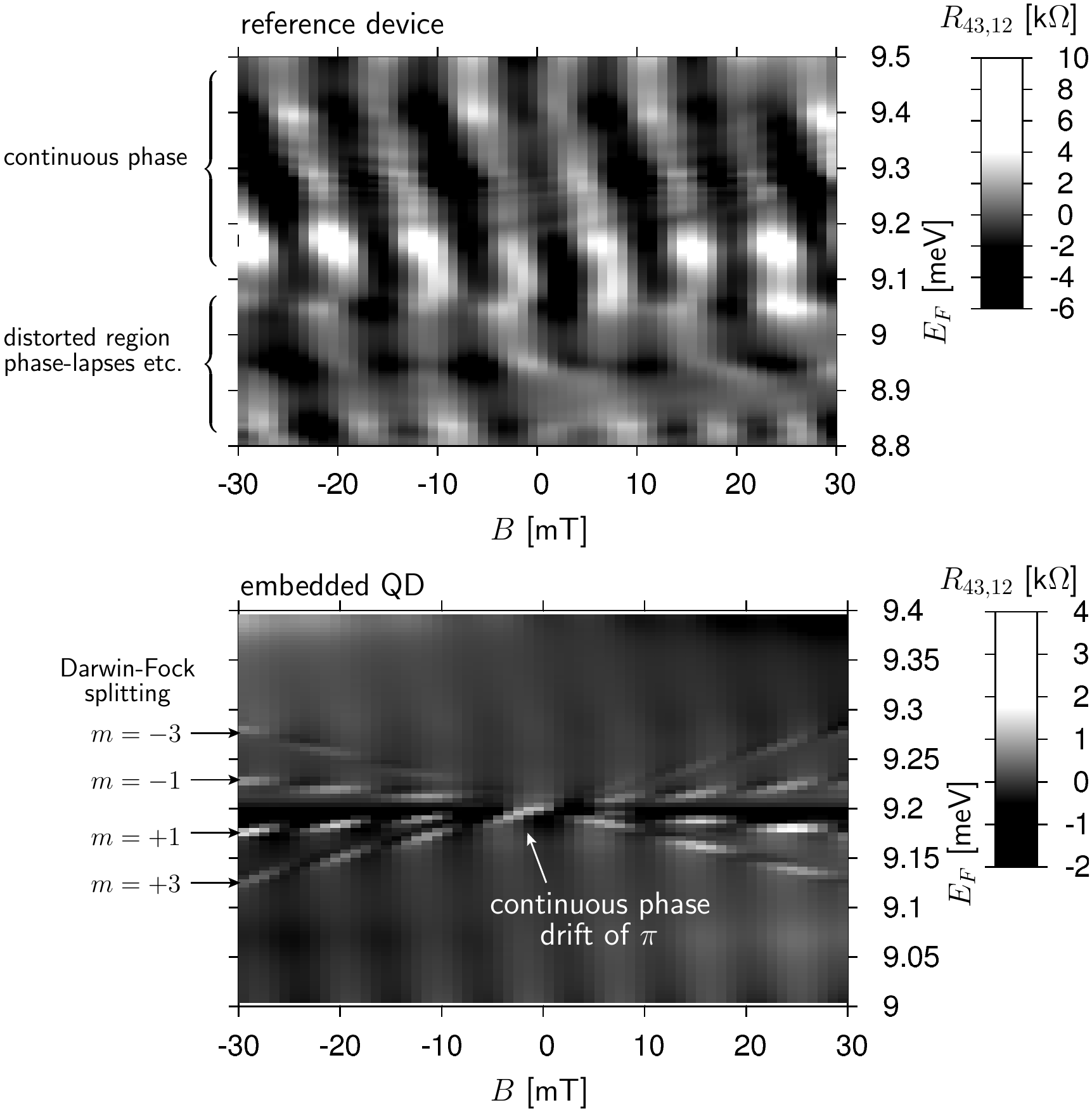}
\caption{\label{fig:CatHairs}
Gray-scale plot of the oscillatory component of the resistance $R_{43,12}$ as a function
of the magnetic field and the Fermi energy.
Upper panel: AB interference fringes for the reference device (4-terminal ring without embedded quantum dot). 
Lower panel: AB interferometer configured as transmission phase detector.
The potential $V_g=-5.8$~meV is adjusted in such a way that the quantum number $n=3$ of the Darwin-Fock spectrum is around $E_F=9.2$~meV, a region for which
the reference device shows a clean phase-drift.
The Darwin-Fock spectrum induces a splitting of the angular momentum states with an increasing
magnetic field and induces {\it cat's} {\it whisker} like interference fringes.
}
\end{center}
\end{figure}
Since already the empty device shows complicated effects \cite{Buchholz2009a, Kreisbeck10, Kobayashi2002a}, a suitable and careful preparation
of the setup is required. Reflection in the junctions and standing waves in the arms lead to resonances and non-trivial scattering effects.
Those show for example as phase lapses in the AB interference fringes \cite{Kreisbeck10}. 
In our simulations (linear regime), see upper panel in Fig.~\ref{fig:CatHairs}, there is a distorted
region in the interference pattern for Fermi energies in the range between 8.8~meV and 9.1~meV.
Therefore, we carefully adjust the range of the Fermi energy under consideration.
There is a well-defined continuous phase drift for Fermi energies in the range between $9.1$~meV to $9.5$~meV. 
Experimentally, the relevant Fermi energy range can be adjusted by
applying a local gate voltage \cite{Schuster97,Avinun05} to shift the energy-levels of the quantum dot relative to the leads and the AB ring.

\section{Darwin-Fock spectrum}
First we consider a two-terminal configuration and calculate the transmission of the harmonic quantum dot embedded in the waveguide $s_2$.
The potential of the harmonic quantum dot is given by (up to some energy, see bottom panel of Fig.~\ref{fig:PotentialABRingembeddedQDot} )
\begin{equation}
 V_{\rm dot}=\frac{1}{2}m^\ast\omega_{\rm dot}^2 (x^2+y^2)+V_g,
\end{equation}
where $\hbar\omega_{\rm dot}$ is the harmonic frequency and $V_g$ models the influence of the local gate-voltage. 
The energy levels of the harmonic quantum dot are given by the Darwin-Fock spectrum \cite{Ferry1997a}
\begin{equation}\label{eq:DarwinFockSpectrum}
 E_{n,m}=(n+1)\hbar\Omega+\frac{1}{2}\hbar\omega_c m+V_g,\quad
\mbox{with} \quad \Omega=\sqrt{\omega^2+\frac{\omega_c^2}{4}}, 
\end{equation}
with $\omega_c=eB/m^\ast$. For the weak applied magnetic fields we neglect the Zeeman spin splitting. 
The main quantum number is defined by $n$, while $m$ denotes
the $n+1$ angular momentum states with $m=-n,\, -n+2,\ ...,\ n$.
The eigenstates in polar coordinates read
\begin{equation}\label{eq:DarwinFockStates}
 \Phi_{n_r,m}(\mbox{\textbf r})=\frac{1}{\sqrt{2\pi}}e^{i m \phi}\frac{1}{l_0}\sqrt{\frac{n_r!}{(n_r+|m|)!}}\,e^{-
r^2/4l_0^2}\,\Big(\frac{r}{\sqrt{2}l_0}\Big)^{|m|}
L_{n_r}^{|m|}\big(r^2/2l_0^2\big),
\end{equation}
with radial quantum number $n_r=(n-|m|)/2$ and $l_0=\sqrt{\hbar/m^\ast\Omega}$. 
$L_{n_r}^{|m|}$ are the generalized Laguerre polynomials. 
\begin{figure}[t!]
\begin{center}
\includegraphics[width=0.99\columnwidth]{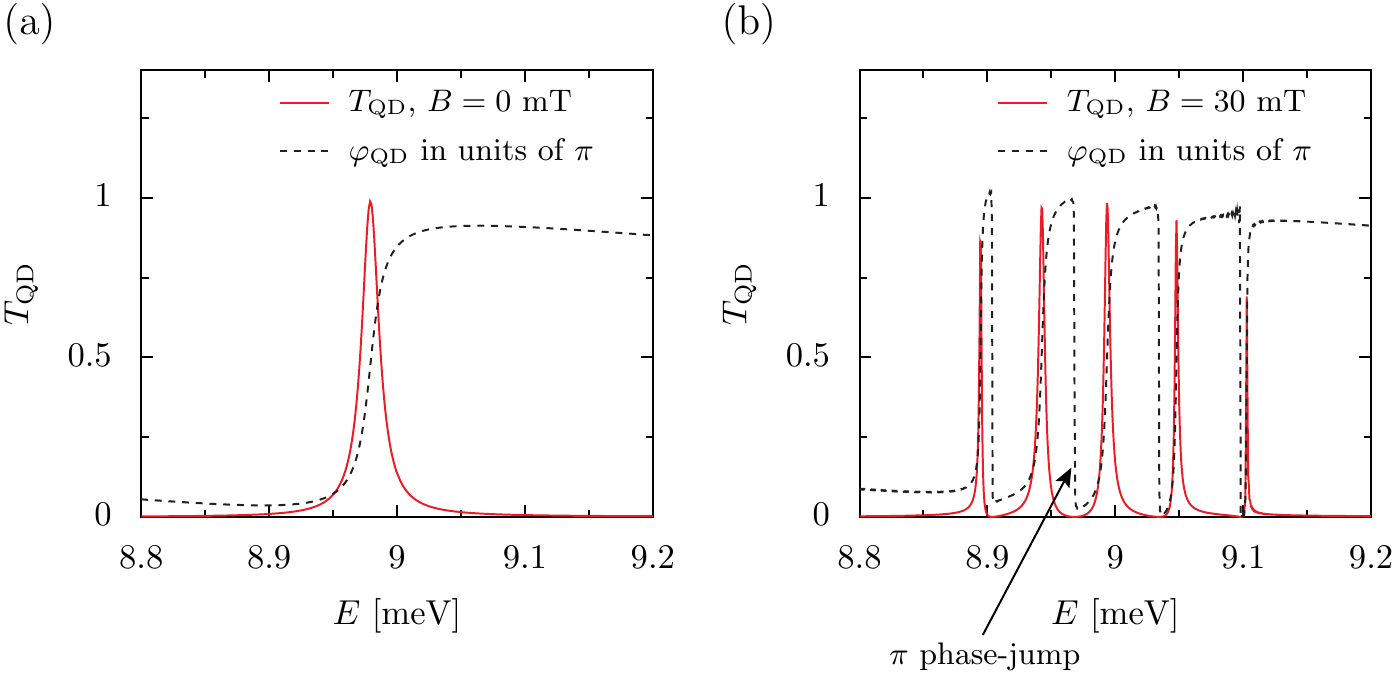}
\caption{\label{fig:DarwinFock_Transmission}
Transmission amplitude $T_{\rm QD}$=$|t_{\rm QD}e^{i \varphi_{\rm QD}}|^2$ as 
function of the Fermi energy in a two-terminal resonant tunneling setup. The dashed line
corresponds to the transmission phase $\varphi_{\rm QD}$ in units of $\pi$. 
Shown is the corresponding Darwin-Fock spectrum for the quantum number $n=4$ ($V_g=-6$~meV, $\hbar\omega_{\rm dot}=3$~meV)
for (a) $B=0$~mT
and (b) $B=30$~mT.
}
\end{center}
\end{figure}

Fig.~\ref{fig:DarwinFock_Transmission} charts the transmission probability
$T_{\rm QD}$=$|t_{\rm QD}e^{i \varphi_{\rm QD}}|^2$ and the transmission phase $\varphi_{\rm QD}$
as functions of the Fermi energy $E_F$.
Without the magnetic field, $B=0$~mT (Fig.~\ref{fig:DarwinFock_Transmission}(a)), the
angular momentum states $m$ of the harmonic oscillator are degenerate.
The transmission probability $T_{\rm QD}$ for transport through
the quantum dot shows a Breit-Wigner resonance around Fermi energy $E_F=9$~meV, corresponding
to the oscillator state with quantum number $n=4$ ($V_g=-6~$~meV, $\hbar\omega_{\rm dot}=3$~meV). 
There is a slight shift of the resonance to lower energies due to the influence of the tunneling junctions.
The transmission phase $\phi_{\rm QD}$ changes continuously by $\pi$, while
scanning the Fermi energy $E_F$ through the resonance. This is expected from
the Breit-Wigner shape and is consistent with the Friedel
sum rule. 
The width of the resonance depends on the strength of the tunneling coupling.

The situation is different for a finite magnetic field. Here, the degeneracy of the
angular momentum $m$ is lifted, and the single resonance
splits into $n+1$ angular states, as is illustrated in Fig.~\ref{fig:DarwinFock_Transmission}(b).
All angular momentum states exhibit the same parity. 
As a consequence the transmission
probability $T_{\rm QD}$ drops to zero between each resonance. Thus, the
continuous
phase drift of $\pi$ at each resonance is interrupted by an abrupt $\pi$
phase-jump. Note that in very high magnetic fields when the different
Zeeman multiplets mix this is no longer the case. In general, it
is expected that zeros in the transmission disappear for finite magnetic
fields and the partial widths of the different states of the dot become complex
numbers \cite{Molina2012a}. However, for harmonic dots connected symmetrically
to the leads this is not the case because of the conservation of parity
in the Darwin-Fock eigenstates of the open dot.

A more realistic description of the electronic states in the dot should include charging energy effects
that lead to Coulomb blockade peaks. A simple but very succesful model for including these effects it the
constant interaction (CI) model. Within the constant interaction model the peaks are separated by the charging
energy of the dot $e^/C$ where $C$ is the dot's capacitance. The modifications of the Darwin-Fock spectrum in the CI model
are well known \cite{Kouwenhoven01} and can describe experimental data in a wide range of parameters \cite{Tarucha96}. Parity properties of
the states are not modified in the CI model and phase behavior should be identical to our description in the following section.
The corrections induced by the CI model to the Darwin-Fock spectrum depend on the ration between charging energy and frequency
of the harmonic potential. The typical values of the charging energy for the geometry we have used in GaAs quantum dots is
$0.3$~meV which is an order the magnitude smaller than the frequency of the potential $\hbar\omega_{\rm dot}=3$~meV \cite{Kouwenhoven97}. For ratios similar to this, the measured results in spectroscopic experiments coincide with the expected from the Darwin-Fock spectrum with only small deviations due to interactions or asymmetries in the dot \cite{Babinski06a,Babinsky06b}.

\section{Zeeman transport spectroscopy}

The transmission phase information is encoded in the interference
fringes of the AB oscillations of the four-terminal measurement. 
We place the quantum dot in arm $s_2$ and run the simulations for the potential landscape
depicted in Fig.~\ref{fig:PotentialABRingembeddedQDot}(b). We adjust $V_g$ such that the resonance $n=3$ of the quantum dot
is located in a region for which the empty device shows a clean phase drift. Here we set $V_g=-5.8$~meV.
We evaluate the non-local resistance $R_{43,12}$ in the linear regime (see \ref{nonlocallinear}).
The lower panel in Fig.~\ref{fig:CatHairs} depicts the oscillatory component of $R_{43,12}$ as function of the magnetic field
and the Fermi energy $E_F$. 
The AB interference emerges when the Fermi energy fulfills the condition for resonant tunneling. 
Only in this situation, both arms contribute to transfer. 
Since the lowest transversal mode carries sufficient kinetic energy to break the tunneling regime there are minor AB oscillations
for off-resonant regions. This is a specific feature of the harmonically capped wave-guides for which the tunneling barrier closes with increasing longitudinal energy.

The splitting of the angular momentum states $m$ with increasing
magnetic field $B$ is nicely visible in the oscillatory component of
$R_{43,12}(B)$ as a function of the Fermi energy. In the AB interference
fringes there emerge four lines reflecting the angular momentum states.
The zeros in the transmission amplitude and the accompanied phase jumps of
$\pi$ create the unique {\it cat's} {\it whisker} like pattern in the non-local resistance $R_{43,12}$.
For small magnetic fields $|B|\approx5$~mT, the resonances of the angular
momentum overlap with each other to form a single broad resonance
that shows a continuous phase drift of~$\pi$.

\section{Conclusions}

For a realistic AB electron interferometer we have performed time-dependent wave-packet
simulations to analyze under which conditions the device can be used as for
measurements of the transmission phase of an embedded quantum dot. Controlling
the properties of the device we find certain
range of parameters that avoid scattering in the junctions and arms of the
interferometer allowing the use of the device for phase measurement.
We have studied an example of a harmonic quantum dot in the presence of magnetic
field showing a Darwin-Fock-Darwin spectrum. The pattern of conductance oscillations
as a function of the magnetic field shows characteristic Zeeman cat's whiskers
that can be used experimentally for simultaneous transport and phase
spectroscopy. In the Zeeman regime, 
transmission zeros and phase lapses appear in between the different resonances
forming the same multiplet as a signature of the same parity of the
corresponding Darwin-Fock eigenstates. This is due to the high symmetry of
these states and in contrast to the general behavior of the phase lapses
that are expected to disappear in the presence of a magnetic field. 
We believe that our realistic simulations are an important contribution to the understanding of the
conditions under which AB interferometers can be used as transport and phase
measurement devices. Similar simulations could serve as a guidance for the
calibration of such a device. Experimental deviations from the simulation results could be used as experimental
signatures of interaction effects.

\section{Acknowledgments}

We would like to acknowledge C. B\"auerle for calling our attention to Refs. \cite{Takada14} and \cite{Takada16} and for useful comments on the first
version of the manuscript.
We acknowledge support through Spanish grant MINECO/FEDER FIS2015-63770-P, CSIC I-Link0936 grant, and from the Deutsche Forschungsgemeinschaft 
(DFG Kr2889).
We thank Nvidia for support via the Harvard CUDA Center of Excellence. 
Part of the computations in this paper were run on the Odyssey cluster, 
supported by the FAS Division of Science, Research Computing Group at Harvard University.

\appendix

\section{Wave-packet propagation and transmission amplitudes}
\label{AppA}
In the ballistic transport regime the total system is composed
of a microscopic scattering region (AB ring with embedded QD) that is coupled to macroscopic contacts via semi-infinite leads.
That is the total Hamiltonian 
\begin{equation}\label{Con1}
 \Ham=\sum_{i}\Ham_i^{\rm asym}+\Ham^{\rm scat}
\end{equation}
splits into asymptotic parts $\Ham_i^{\rm asym}$, describing the leads and a scattering
part $\Ham^{\rm scat}$. The leads are semi-infinite quasi one-dimensional objects that are quantized in transversal direction.
We use a time-dependent approach based on wave packet propagation \cite{Kramer10, Garashchuk1998a, Tannor2000a} to
compute the transmission amplitude $t_{i n_i, j n_j}$, which corresponds to the probability
amplitude of scattering from lead~$j$ populating transversal mode~$n_j$ to lead~$i$ populating
transversal mode~$n_i$. 
Within the Landauer-B\"uttiker formalism, the microscopic
current in each lead is related to the transmission $T_{i,j}=\sum_{n_i,n_j}|t_{i n_i,j n_j}|^2$ by
\begin{equation}\label{ILB}
I_i=\frac{e}{h}\int \mbox{d}E\sum_j T_{i,j}(E)(f_i(E)-f_j(E)).
\end{equation}
Here, $f_i(E)=(\exp(E-\mu_i)/k_BT + 1)^{-1}$ describe the Fermi-Dirac distribution for each lead with chemical potential $\mu_i$.

The scattering eigenstates of the full Hamiltonian can be identified with the
corresponding eigenstates of the asymptotic leads 
\begin{equation}\label{Con2}
\Ham_i^{\rm asym}\ \psi_{i,n_i,\pm}(E)=E\ \psi_{i,n_i,\pm}(E)
\end{equation}
with
\begin{equation}\label{Con3}
 E=\frac{\hbar^2 k_i^2}{2 m^\ast}+E_{i,n_i,\pm k_i}
\end{equation}
and
\begin{equation}\label{Con4}
 \psi_{i,n_i,\pm}(E)=\left(\frac{\partial E}{\partial k_i}\right)^{-\frac{1}{2}}\chi_{i,n_i,\pm k_i}\,
e^{\pm\ims\, k_i x_i}.
\end{equation}
Here, $\chi_{i,n_i,\pm k_i}$ is the $n_i$-th transversal mode of lead~$i$ with transversal
energy $E_{i,n_i,\pm k_i}$. The remaining energy goes into the longitudinal
kinetic energy part $E_{\rm kin}=\hbar^2 k_i^2/2 m^\ast$. 
The longitudinal orientation of the leads points along coordinate $\hat{x}_i$. Positive
longitudinal wave vectors $k_i$ correspond
to incoming waves. 
In the presence of a magnetic field, the transversal modes depend on the
longitudinal momentum.

In our simulations the eigenstates in the asymptotic leads $\chi_{i,n_i,\pm k_i}\,e^{\pm\ims\, k_i x_i}$ satisfy the normalization condition 
$\langle \chi_{i,n_i,\pm k_i}\,e^{\pm\ims\, k_i x_i}|\chi_{i,n_i',\pm k_i'}\,e^{\pm\ims\, k_i' x_i}\rangle=\delta(k_i-k'_i) \delta_{n_i,n_i',\pm,\pm}$.
Therefore, we need to introduce the factor $\left(\partial E/\partial k_i\right)^{-\frac{1}{2}}$ in eq.~(\ref{Con4}) to ensures the normalization condition in 'energy space'

\begin{equation}\label{Con5}
 \scalarprod{\psi_{i,n,\pm}(E)}{\psi_{i,n',\pm}(E')}=\delta_{nn',\pm\pm}\delta(E-E').
\end{equation}
The asymptotic eigenfunctions form a complete orthonormal set in lead~$i$.
\begin{figure}[t!]
\begin{center}
\includegraphics[width=0.8\columnwidth]{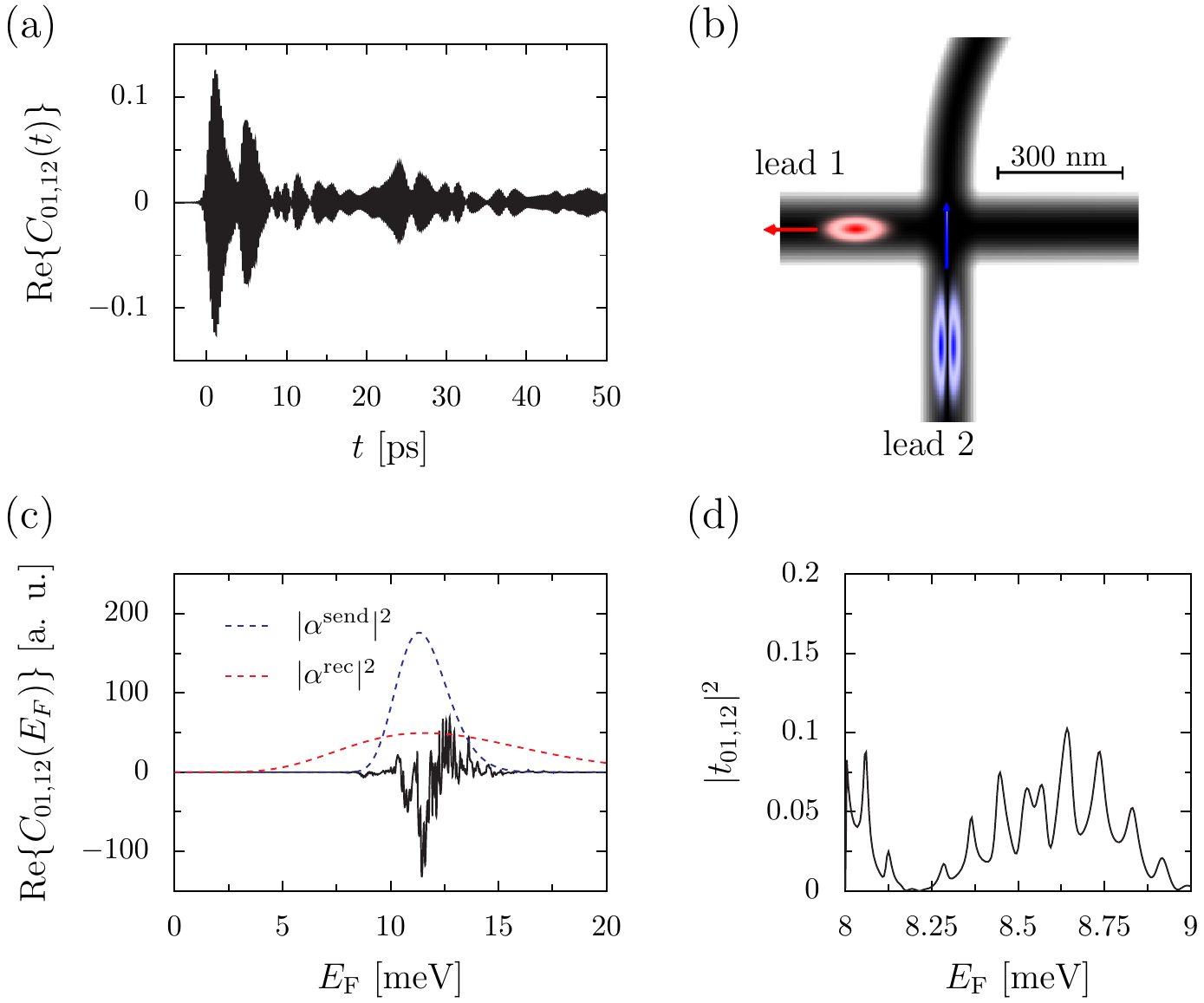}
\caption{\label{fig:ACT_ACE}
Illustration of the steps involved in the wave-packet approach to
get the transmission probability $|t_{01,12}|^2$ for the device geometry shown in Fig.~\ref{fig:PotentialABRingembeddedQDot} in the main text. (a) time correlation
function. (b) visualizes the spatial extension of the receiver (red) and sender (t=0~ps) (blue) wave-packet.
(c) real part of the energy representation of the cross-correlation function in arbitrary units.
The dashed lines correspond to the energy density of the receiver $|\alpha^{\rm rec}|^2$ (red)
and sender $|\alpha^{\rm send}|^2$ (blue) wave 
packet. (d) transmission probability $|t_{01,12}|^2$ calculated according to eq.~(\ref{Con14}).}
\end{center}
\end{figure}

Scattering eigenstates of the total Hamiltonian are solutions
of the Schr\"o\-ding\-er equation that is restricted to the
asymptotic regions. 
We define 
\begin{equation}\label{Con6}
 \Ham\, \psi^\pm_{i,n_i}(E)=\left(\frac{\hbar^2 k_i^2}{2 m^\ast}+E_{i,n_i,\pm k_i}\right)\, \psi^\pm_{i,n_i}(E)=E\,
\psi^\pm_{i,n_i}(E),
\end{equation}
where $\psi^\pm_{i,n_i}(E)$ is related to the asymptotic
eigenstates $\psi_{i,n_i, \pm}(E)$ of lead $i$. 
In the corresponding time picture, $\psi_{i,n}^+$ starts with an incoming wave
in lead $i$ with fixed
momentum $k_i$ populating transversal mode~$n_i$. After the wave
hits the scattering region
it gets either reflected 
in lead $i$ or transmitted to outgoing waves in other asymptotic leads $j\ne i$.
Due to inter-mode scattering, the reflected and transmitted parts are superpositions of
different transversal modes.
The scattering eigenstates $\psi^-_{i,n_i}$ correspond to the time-reversed situation
\begin{equation}\label{Con7}
\psi^-_{i,n_i}(B)=\left(\psi^+_{i,n_i}(-B)\right)^\ast.
\end{equation}
All scattered parts unify in a single outgoing wave with wave vector $k_i$ and transversal mode $n_i$.
$\psi^+_{i,n}$ describes the history, whereas $\psi^-_{i,n}$ describes the
destiny of the states. This gives an intuitive picture of the definition of the scattering matrix
\begin{equation}\label{Con8}
 t_{i n_i, j n_j}(E)\delta(E-E')=\scalarprod{\psi^-_{i,n_i}(E)}{\psi^+_{j,n_j}(E')},
\end{equation}
which can be interpreted as the projection of the \textit{destiny} on the \textit{history}. 

In the following, we establish the concepts behind the wave-packet approach. 
Let us define an incoming sender wave packet which is
is located in one of the asymptotic leads where it populates a specific
transversal mode. Moreover, it is composed
of purely incoming longitudinal momenta. 
Thus the sender wave-packet is represented in terms of
scattering eigenstates $\psi^+_{j,n}(E)$, 
\begin{equation}\label{Con10}
 \phi_{j,n_j}^{\rm send}=\int \mbox{d}E\ \alpha^{\rm send}_{j,n_j}(E)\, \psi_{j,n_j}^+(E).
\end{equation}
In an analogous manner, we define a receiver wave-packet that is located in 
lead~$i$ where it populates transversal mode~$n_i$, and is composed of purely outgoing
longitudinal momenta,
\begin{equation}\label{Con11}
 \phi_{i,n_i}^{\rm rec}=\int\mbox{d}E\ \alpha_{i,n_i}^{\rm rec}(E)\, \psi_{i,n_i}^-(E).
\end{equation}
The receiver represents
the destiny of the scattered wave and can be interpreted as a detector.
The scattering eigenstates are recovered by 
\begin{eqnarray}\label{Con12}
 \psi_{i,n_i}^{\pm}(E) =\frac{(2\pi\hbar)^{-1}}{\alpha^{\rm send/rec}_{i,n_i}(E)}\int_{-\infty}^\infty\mbox{d}t\ 
\phi_{i,n_i}^{\rm send/rec}(t)\ e^{\ims Et/\hbar},
\end{eqnarray}
where
\begin{equation}\label{Con13}
 \phi_{i,n}^{\rm send/rec}(t)=e^{-\ims \Ham t/\hbar}\ \phi_{i,n}^{\rm send/rec}(t_0=0)
\end{equation}
corresponds to the time evolution of the wave packets. During the propagation of the 
wave packet, it successively traverses the potential landscape and its
time-correlation function recovers the stationary solutions of the
underlying Hamiltonian.
We insert eq.~(\ref{Con12}) into the definition of 
the scattering matrix eq.~(\ref{Con8}) and recover the transmission amplitudes
\begin{eqnarray}\label{Con14}
 t_{i n_i,j n_j}(E)=\frac{(2\pi\hbar)^{-1}}{(\alpha^{\rm rec}_{i,n_i}(E))^{^\ast}\,
\alpha^{\rm send}_{j,n_j}(E)}\int_{-\infty}^\infty \hspace{-0.4cm}\mbox{d}t\ 
C_{i n_i,j n_j}(t)e^{\ims Et/\hbar}
\end{eqnarray}
in terms of the cross-correlation function
\begin{eqnarray}\label{Con15}
C_{i n_i,j n_j}(t)=\scalarprod{\phi_{i,n_i}^{\rm rec}|e^{\ims \Ham t/\hbar}}{\phi_{j,n_j}^{\rm send}}
=\scalarprod{\phi_{i,n_i}^{\rm rec}}{\phi_{j,n_j}^{\rm send}(t)}.
\end{eqnarray}
The factors $\alpha^{\rm send/rec}_{i,n}$ depend on the specific shape of the used wave packets
\begin{eqnarray}\label{Con16}
 \alpha^{\rm send/rec}_{i,n_i}(E)=\scalarprod{ \psi_{i,n_i}^{+/-}(E)}{\phi_{i,n_i}^{\rm send/rec}}.
\end{eqnarray}
Since the wave packets $\phi_{i,n_i}^{\rm send/rec}$ are located in the asymptotic
channels, 
the scattering eigenstates are associated with
lead eigenstates. Thus eq.~(\ref{Con16})
reduces to 
\begin{eqnarray}\label{Con17}
 \alpha^{\rm send/rec}_{i,n_i}(E)=\left(\frac{\partial E}{\partial k_i}\right)^{-\frac{1}{2}}\langle
\chi_{i,n_i,\pm k_i}\,e^{\pm\ims\, k_i x_i}|\phi_{i,n_i}^{\rm send/rec}\rangle.
\end{eqnarray}

\section{Wave-packet evolution in the embedded dot configuration}
\label{AppB}
In the following we apply the wave-package approach to the four-terminal AB ring with the embedded quantum dot.
As illustrative example, we compute the transmission amplitude $t_{10,21}(E)$ for inter-mode scattering from
transversal mode~$1$ in lead~$2$ to transversal mode~$0$ in lead~$1$. 
The corresponding receiver and sender wave-packets are illustrated in Fig.~\ref{fig:ACT_ACE}(b).
During the propagation of the sender wave-packet, we keep track of
the cross-correlation function $C_{10,21}(t)$ defined in 
eq.~(\ref{Con15}).
The cross-correlation function reflects the choreography of the wave packet and shows rich structures such as revivals in amplitude, see Fig.~\ref{fig:ACT_ACE}(a).
We perform the Fourier-transform (Fig.~\ref{fig:ACT_ACE}(c)) and follow eq.~(\ref{Con14}) to obtain the transmission probability $|t_{10,21}(E)|^2$ depicted in
Fig.~\ref{fig:ACT_ACE}(d).

\begin{figure}[t!]
\begin{center}
\includegraphics[width=0.9\columnwidth]{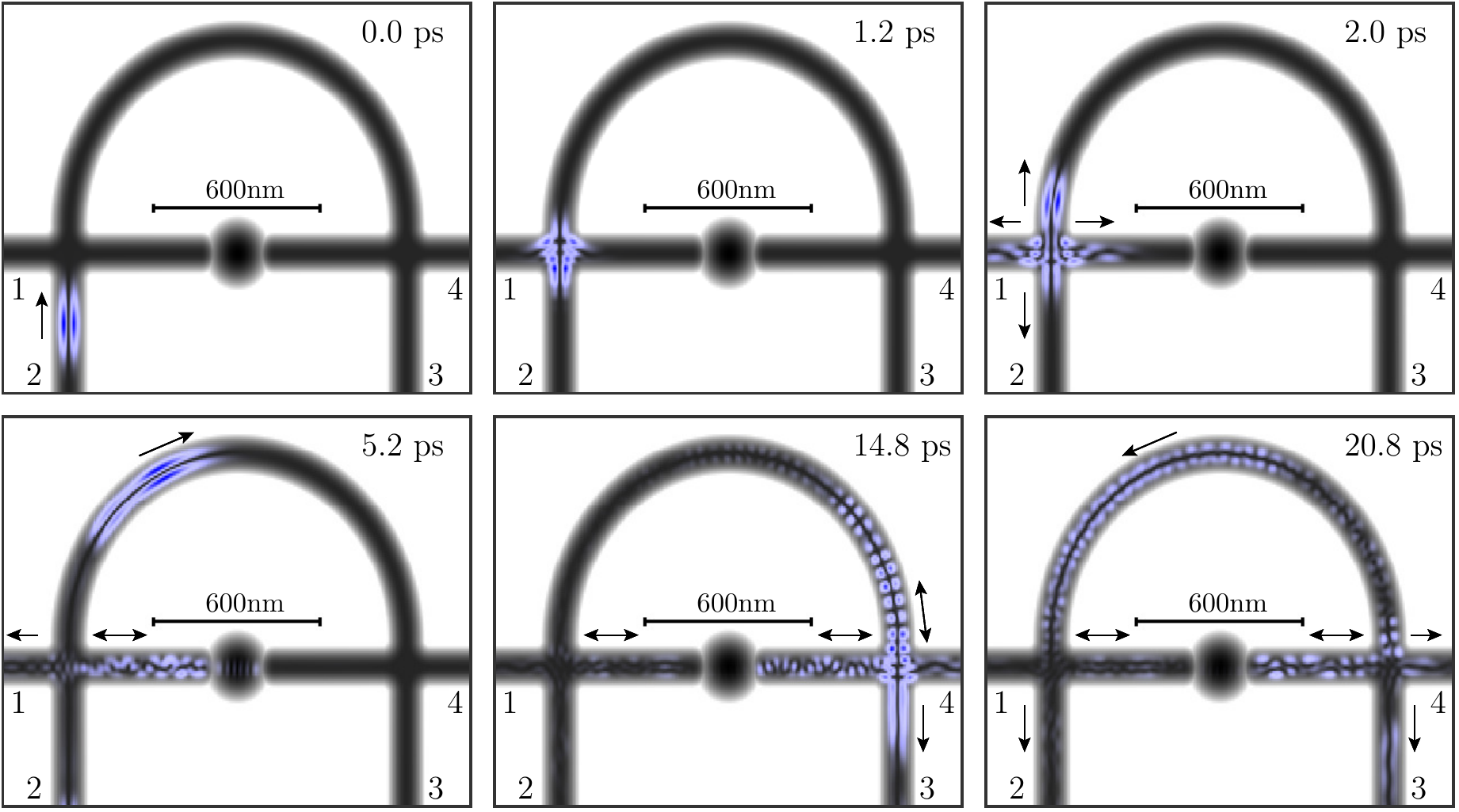}
\caption{\label{fig:Snapshots}
Snapshots of the wave packet propagation. The color (blue) encodes the amplitude of the wave packet. During
the time evolution, the wave undergoes multiple scattering events. The arrows indicate the motion of the scattered parts.}
\end{center}
\end{figure}

Fig.~\ref{fig:Snapshots} shows snapshots of the time evolution of the sender
wave-packet. The arrows indicate the motion of the wave packet components.
The propagation is done with the split-operator method \cite{Feit1982a}.
Initially (0.0~ps), the sender is located in the asymptotic lead~2 and
propagates toward the left cross-junction. At 1.2~ps, the sender
approaches the center of the junction, scatters in different directions and splits into four parts as is illustrated
in the snapshot at 2.0~ps. Some parts get reflected
into an outgoing wave in lead~2. Other parts scatter to lead~1 and are detected
by the receiver wave-packet. This leads to the large amplitude in
$C_{01,12}(t)$ (see Fig.~\ref{fig:ACT_ACE}(a)) for the time interval between 1~ps and 3~ps.
The remaining parts scatter either to the upper arm $s_1$ or to the lower arm $s_2$ of the ring.
These parts travel along the ring toward the right cross-junction.
The lower arm $s_1$ contains the embedded quantum dot and only few parts of the
wave packet that fulfill the conditions for resonant tunneling are transmitted to the right cross-junction. The parts that get reflected
at the QD undergo multiple scattering at the left part of the lower arm (see snapshot 5.2~ps), and finally are transferred to lead~1. This process is 
responsible for the first revival seen in the cross-correlation function around 5~ps~\---~7~ps.
At about 14.8 ps the dominant part of the wave packet traveling along the upper arm of the ring reaches the right-cross junction.
Some parts get reflected and go back along the upper arm until they reach again the left junction at around 20.8~ps.

\section{Evaluation of the non-local resistance in the linear transport regime}\label{nonlocallinear}
According to the Landauer-B\"uttiker formalism \cite{Landauer70,Buttiker86}, the microscopic currents in each lead of the four terminal AB interferometer are given by eq.~(\ref{ILB}).
In the following we consider the linear transport regime, for which we assume that the transport window opened by the difference in the Fermi functions in eq.~(\ref{ILB}) is smaller than the energy scale of fluctuations in the transmission amplitudes $t_{in_i,jn_j}(E)$ (obtained from eq.~(\ref{Con14})). The microscopic currents are then expressed by linear equations  
\begin{equation}
  I_i = \frac{e}{h}\Big((1-T_{ii}(E_F)\mu_i-\sum_{j\neq i}T_{ij}(E_F)\mu_j\Big), \mbox{with }T_{i,j}=\sum_{n_i,n_j}|t_{i n_i,j n_j}|^2
\end{equation}
and the local resistance $R_{43,12}$ can be evaluated analytically \cite{Buttiker86}
\begin{equation}
 R_{43,12}(E_F) = \frac{h}{e^2}\frac{T_{14}(E_F)T_{23}(E_F)-T_{13}(E_F)T_{24}(E_F)}{D(E_F)}
\end{equation}
with
\begin{eqnarray}
  \hspace{-0.8cm}D(E) &=&  \frac{h^2}{e^4}(\alpha_{11}(E)\alpha_{22}(E)-\alpha_{12}(E)\alpha_{21}(E))S(E)\nonumber\\
  \hspace{-0.8cm}\alpha_{11}(E)&=&\frac{e^2}{hS(E)}\Big((1-T_{11}(E))S(E)-(T_{14}(E)+T_{12}(E))(T_{41}(E)+T_{21}(E))\Big)\nonumber\\
  \hspace{-0.8cm}\alpha_{22}(E)&=&\frac{e^2}{hS(E)}\Big((1-T_{22}(E))S(E)-(T_{21}(E)+T_{23}(E))(T_{32}(E)+T_{12}(E))\Big)\nonumber\\
  \hspace{-0.8cm}\alpha_{12}(E)&=&\frac{e^2}{hS(E)}\Big((T_{12}(E)T_{34}(E)-T_{14}(E)T_{32}(E))\Big)\nonumber\\
  \hspace{-0.8cm} \alpha_{21}(E)&=&\frac{e^2}{hS(E)}\Big((T_{21}(E)T_{43}(E)-T_{41}(E)T_{23}(E))\Big)\nonumber\\
  \hspace{-0.8cm} S(E)&=&T_{12}(E)+T_{14}(E)+T_{32}(E)+T_{34}(E).
\end{eqnarray}


\section*{References}

\begin{thebibliography}{100}

\bibitem{Landauer70} Landauer R 1970 {\it Philos. Mag.} \textbf{21} 863\-7

\bibitem{Buttiker86} B\"uttiker M 1986 {\it Phys. Rev. Lett.} \textbf{57} 1761\-4

\bibitem{Yacoby95}
Yacoby A, Heiblum M, Mahalu D, Umansky V and Shtrikman H 1995 {\it Phys. Rev. Lett.} \textbf{74} 4047\--50

\bibitem{Onsager31}
Onsager L 1931 {\it Phys. Rev.} \textbf{38}, 2265\--79

\bibitem{Schuster97}
Schuster R, Buks E, Heiblum M, Mahalu D, Umansky V and
Shtrikman H 1997 {\it Nature} \textbf{385} 417\--20

\bibitem{Sigrist04}
Sigrist M, Fuhrer A, Ihn T, Ensslin K, Ulloa S E, Wegscheider W and Bichler M 2004 \textit{Phys. Rev. Lett.} \textbf{93} 066802

\bibitem{Gerland00}
Gerland U, von Delft J, Costi T A and Oreg Y 2000 {\it Phys. Rev. Lett.} \textbf{84} 3710\--13

\bibitem{Ji00} Yang Ji, Heiblum M, Sprinzak D, Mahalu D and Shtrikman H 2000 {\it Science} \textbf{290} 779\--83

\bibitem{Ji02} Yang Ji, Heiblum M and Shtrikman H 2002 {\it Phys. Rev. Lett.} \textbf{88}, 076601

\bibitem{Zaffalon08}
Zaffalon M, Aveek Bid, Heiblum M, Mahalu D, Umansky V 2008 {\it Phys. Rev. Lett.} \textbf{100} 226601

\bibitem{Takada14}
Takada S, B\"auerle C, Yamamoto M, Watanabe K, Hermelin S, Meunier T, Alex A, Weichselbaum A, von Delft J, Ludwig A, Wieck A D, Tarucha S 2014 {\it Phys. Rev. Lett.} \textbf{113} 126601

\bibitem{Takada16} Takada S, Yamamoto M, B\"auerle C, Alex A, von Delft J, Ludwig A, Wieck A D, Tarucha S 2016 {\it Phys. Rev. B} \textbf{94}, 081303(R)

\bibitem{Avinun05}
Avinun-Kalish M, Heiblum M, Zarchin O, Mahalu D and Umansky V 2005
{\it Nature} \textbf{436} 529\--33


\bibitem{LevyYeyati95} 
Yeyati A L and B\"uttiker M 1995 {\it Phys. Rev. B} \textbf{52} R14360

\bibitem{Lee99} Lee H-W 1999 {\it Phys. Rev. Lett.} \textbf{82}, 2358\--61

\bibitem{Taniguchi99} Taniguchi T and B\"uttiker M 1999 {\it Phys. Rev. B} \textbf{60} 13814\--23

\bibitem{LevyYeyati00}
Yeyati A L and B\"uttiker M 2000 {\it Phys. Rev. B} \textbf{62} 7307\--15

\bibitem{Hackenbroich97}
Hackenbroich G, Heiss W D and Weidenm\"uller H A 1997 {\it Phys. Rev. Lett.} \textbf{79} 127\--30

\bibitem{Baltin99a}
Baltin R, Gefen Y, Hackenbroich G and Weidenm\"uller H A 1999 {\it Eur. Phys. J. B} \textbf{10} 119\--29

\bibitem{Silvestrov00}
Silvestrov P G and Imry Y 2000 {\it Phys. Rev. Lett.} \textbf{85} 2565\-8

\bibitem{Silvestrov07}
Silvestrov  P G and Imry Y 2007 {\it New J. Phys.} \textbf{9} 125

\bibitem{Goldstein09}
Goldstein M, Berkovits R, Gefen Y and Weidenm\"uller H A 2009 {\it Phys. Rev. B} \textbf{79} 125307

\bibitem{Karrasch07}
 Karrasch C, Hecht T, Weichselbaum A, Oreg Y, von Delft F and Meden V 2007 {\it Phys. Rev. Lett.} \textbf{98} 186802

\bibitem{Silva02}
Silva A, Oreg Y and Gefen Y 2002 {\it Phys. Rev. B} \textbf{66} 195316

\bibitem{Oreg07} Oreg Y 2007 {\it New J.\ Phys.} \textbf{9} 122

\bibitem{Molina12}
Molina R A, Jalabert R A, Weinmann D and Jacquod P 2012 {\it Phys. Rev. Lett.} \textbf{108} 076803

\bibitem{Molina13} Molina R A, Schmitteckert P, Weinmann D, Jalabert R A and Jacquod P 2013 {\it Phys. Rev. B} \textbf{88} 045419

\bibitem{Jalabert14} Jalabert R A, Weick G, Weidenm\"uller H A and Weinmann D 2014 \textit{Phys. Rev. E} \textbf{89} 052911

\bibitem{Wu98} Wu J, Gu B-L, Chen H, Duan W and Kawazoe Y 1998  
\textit{Phys. Rev. Lett.} \textbf{80} 1952\-5

\bibitem{Kang99}
Kang K 1999 \textit{Phys. Rev. B} \textbf{59} 4608\--11


\bibitem{Aharony02} Aharony A, Entin-Wohlman O, Halperin B I and Imry Y 2002 {\it Phys. Rev. B} {\bf 66} 115311

\bibitem{EntinWohlman02}
Entin-Wohlman O, Aharony A, Imry Y, Levinson Y and Schiller A 2002 
{\it Phys. Rev. Lett.} {\bf 88} 166801

\bibitem{Takada15} Takada S, Yamamoto M, B\"auerle C, Watanabe K, Ludwig A, Wieck A D and Tarucha S 2015
{\it Appl. Phys. Lett.} \textbf{107} 063101

\bibitem{Buchholz2009a} Buchholz S S, Fischer S F, Kunze U, Reuter D and Wieck A D 2009 {\it Appl. Phys. Lett.} \textbf{94} 022107

\bibitem{Kreisbeck10} Kreisbeck C, Kramer T, Buchholz S S, Fischer S F, Kunze U,
Reuter D and Wieck A D 2010 {\it Phys. Rev. B} \textbf{82} 165329

\bibitem{Kramer08} 
Kramer T, Heller EJ and Parrott RE 2008 {\it J. Phys. Conf. Series.} \textbf{99} 012010

\bibitem{Chaves09}
Chaves A, Farias GA, Peeters FM and Szafran B 2009 {\it Phys. Rev. B} \textbf{80} 125331

\bibitem{Kramer10}
Kramer T, Kreisbeck C and Krueckl V 2010 {\it Phys. Scr.} \textbf{82} 038101

\bibitem{Kramer16}
Kramer T, Kreisbeck C, Riha C, Chiatti O, Buchholz S S, Wieck A D, Reuter D and Fischer~S~F 2016 {\it AIP Advances} \textbf{6} 065306

\bibitem{Buchholz11}
Buchholz S S, Kunze U, Reuter D, Wieck A D and Fischer S F 2011
{\it App. Phys. Lett.} \textbf{98} 102111 

\bibitem{Baranger91} Baranger H U, DiVicenzo D P, Jalabert R A and Stone A D 1991 {\it Phys. Rev. B} \textbf{44} 10637\--75

\bibitem{Kobayashi2002a}
 Kobayashi K, Aikawa H, Katsumoto S and Iye Y 2002 {\it J. Phys. Soc. Jpn.} \textbf{71} 2094\-7

\bibitem{Ferry1997a} Ferry D K and Goodnick S M 1997 {\it Transport in
    nanostructures} (Cambridge University Press)

\bibitem{Molina2012a}
 Molina R A, Schmitteckert P, Weinmann D, Jalabert R A, Jacquod P 2012 {\it J. Phys.: Conf. Ser.} \textbf{338} 012011

\bibitem{Kouwenhoven01} Kouwenhoven L P, Austing D G, Tarucha S 2001 {\it Rep. Prog. Phys.} \textbf{64} 701\--36

\bibitem{Tarucha96} Tarucha S, Austing D G, Honda T, van der Hage R J, Kouwenhoven L P 1996 {\it Phys. Rev. Lett.} \textbf{77} 3613/-6

\bibitem{Kouwenhoven97} Kouwenhoven L P, Marcus C M, McEuen P L, Tarucha S, Westervelt R M, Wingreen N S 1997 {\it Electron transport in quantum dots} on {\it Mesoscopic electron transport} (Kluwer)

\bibitem{Babinski06a} Babinski A, Potemski M, Raymond S, Lapointe J, Wasilewski Z R 2006 {\it Phys. Rev. B} \textbf{74} 155301 
    
\bibitem{Garashchuk1998a} Garashchuk S and Tannor D J 1998 {\it J. Chem. Phys.} {\bf 109} 3028\--36

\bibitem{Tannor2000a} Tannor D J and Garashchuk S 2000 {\it Annu. Rev. Phys. Chem.} {\bf 51} 553\---600

\bibitem{Feit1982a} Feit M D, Fleck J A and Steiger A 1982 {\it J. Comp. Phys.} \textbf{47} 412\--33


\end{thebibliography}
\end{document}